\documentstyle[aas2pp4,tighten]{article}

\def\kms{kms$^{-1}$}

\lefthead{Moles et al.}
\righthead{Stephan's Quintet}

\begin{document}

\title{The dynamical status of Stephan's Quintet}

\author{M. Moles$^{1,2}$, J. W. Sulentic$^3$, and I. M\'arquez$^{4,5}$}

\altaffiltext{1}{Instituto de Matem\'aticas y F\'{\i}sica Fundamental,
CSIC, C/ Serrano 123, 28006 Madrid, Spain}

\altaffiltext{2}{Observatorio Astron\'omico Nacional, Madrid, Spain}

\altaffiltext{3}{Department of Physics and Astronomy, University of
Alabama, Tuscaloosa, USA 35487}

\altaffiltext{4}{Institut d'Astrophysique de Paris, 98bis bd. Arago,
75014 Paris, France}

\altaffiltext{5}{Instituto de Astrof\'{\i}sica de Andaluc\'{\i}a (CSIC), 
Granada, Spain}



\begin{abstract}

Multiwavelength data for Stephan's Quintet (SQ) are consistent with the
following model for this compact galaxy group.  (1) Discordant redshift
NGC~7320 is an unrelated foreground galaxy.  (2) In the past SQ was an
accordant redshift quartet involving NGC~7317, 18A, 19 and 20C.
NGC~7320C collided (probably not for the first time) with the group a
few times 10$^8$ years ago and stripped the interstellar matter
 from NGC~7319.  (3) In
the present SQ is again an accordant quartet involving NGC~7317, 18A,B,
and 19. NGC~7318B is now entering the group at high velocity for the 
first time, giving rise to a shock zone. If most compact groups are 
like  SQ, then they are frequently visited by infalling neighbors that 
perturb the group and themselves. SQ represents strong evidence for 
secondary infall in a small group environment. Tidal stripping reduces 
the mass of the infalling galaxies, thereby increasing the timescale for 
their orbital decay.  There is little evidence that these high velocity 
``intruders'' are rapidly captured and/or merge with the system.  
Instead they are the mechanism that sustains compact groups against 
collapse. Efficient gas stripping may account for the low star formation 
rate observed in compact groups and infall of residual gas 
into galactic nuclei may also foster the onset of active galactic nucleus
 activity.  
\end{abstract}

\section{Introduction}

Stephan's Quintet (SQ) epitomizes the problems that compact groups pose
for our ideas about galaxy formation and evolution. It is one of the
most luminous and high surface brightness aggregates included in the
first reasonably complete catalog (Hickson 1982; hereafter HCG) of
compact groups (HCG92). Questions center around how many of the
galaxies are in close proximity to one another, as well as how long and
violently they have been interacting. These questions are particularly
relevant because one of the components, NGC~7320, shows a discordant
redshift. Finally, there is the question of whether SQ is representative
of the compact group phenomenon. It is relevant for the latter question
to point out that the accordant redshift part of SQ satisfies the HCG
selection criteria - Stephan's Quartet is also a compact group.

SQ is composed of a  {\sl kernel} of three galaxies (NGC~7317, 18A and
19) with very low velocity dispersion (cz= 6563, 6620 and 6650 \kms~
respectively; see Moles et al. 1997; hereafter paper I).  NGC~7320 (800
\kms) and NGC~7318B (5765 \kms) complete the apparent compact group. Another 
accordant
redshift galaxy (NGC~7320C, cz$\sim$6000 \kms) lies 3 arcmin ENE of
NGC~7319. NGC~7320C is sufficiently faint that its proximity does not
violate the HCG isolation criteria. A much brighter spiral galaxy
NGC~7331 lies 30 arcmin NE of SQ and shows a redshift (cz= 821 \kms)
similar to NGC~7320.  All seven of these galaxies play a role in our
interpretation of the dynamical state of SQ.  Figure 1a shows a
schematic of SQ and environs, while Figure 2a shows a wide band image of
SQ proper. Wide field images can be found in the plates of Arp and
Kormendy (1972).

We discuss new optical observations (presented in paper I), as well as
published X-ray and radio data. We use them to infer a dynamical
history for SQ. Section 2 considers the relationship of
discordant-redshift NGC~7320 and the remaining SQ members. We consider
the past and present dynamical states of the accordant group in
sections 3 and 4 respectively.  In section 5 we consider the
implications of SQ as representative of the compact group phenomenon.

\section{NGC~7320: A Late Type Spiral Projected on an Accordant
Quartet}

Several investigators have pointed out the low probability for
NGC~7320 to be a chance projection on the accordant quartet in SQ
(Burbidge and Burbidge 1961; Arp 1973). Little weight would be given to
such an {\it a posteriori} calculation were it not for the fact that
discordant components are so numerous in compact groups ($\sim$43/100
groups in HCG contain at least one discordant-redshift member; see
Sulentic 1994).  NGC~7320 shows some peculiarities (see paper I) that
could be interpreted as evidence for interaction with the higher
redshift SQ members. The most significant involves a blue tidal tail
that emerges from the SE end of NGC~7320 (Arp and Kormendy 1972). If we
assume that the tidal tail is evidence for recent galaxy-galaxy
interaction, we are left with three possible interpretations:

(1) it involves only the high redshift members of SQ. In this case
NGC~7320 is a foreground galaxy projected on one of the background
manifestations of this interaction; or

(2) it involves a past encounter between NGC~7320 and NGC~7331. In this
case, it is a foreground tail projected near the background high
redshift quartet; or

(3) it involves direct interaction between NGC~7320 and the accordant
SQ. This would require the assumption of non-Doppler redshifts in one
(or four) components because the velocity difference between the
components is too high for a pure dynamical explanation.

The tail emerging from NGC~7320 is parallel to a second narrower and
brighter tail that emerges from one of the the spiral arms in NGC~7319
(see Figure 1b) and extends in the direction of similar redshift
NGC~7320C. This fact clearly favors hypothesis 1 because it relates the
``NGC7320 tail'' to one that is unambiguously related to interaction
involving only higher redshift members. The tail emerging from NGC~7320
can actually be traced even farther than the brighter one, and it
extends directly to NGC~7320C (Arp and Lorre 1976). Ambiguity in the
hypothesis 1 interpretation stems from the fact that NGC~7331 is
located in the same direction, but beyond NGC~7320C.  Further support
for hypothesis 1 comes from a low-redshift H$\alpha$ image (provided by
W. Keel) shown in Figure 2b. The distribution of HII regions in
NGC~7320 is symmetric about its nucleus with no evidence for any low
redshift H$\alpha$ emission in the tidal tail.

Standard redshift-independent distance estimates for NGC~7320 as well
as NGC~7317 and 18A,B  have been attempted (paper 1; see also Kent
1981; Sulentic 1994). It is unclear whether techniques calibrated with
normal galaxies can be reasonably applied to galaxies that are
suspected to be, or are manifestly abnormal. The debate over the
interpretation of the HI data illustrates this point very well
(Sulentic 1994).  On the other hand, we show in paper I that some of
the galaxies in SQ show only small disturbances (including NGC~7320, if
the tail does not belong to it) and, therefore employ normal distance
estimators. Indeed, those standard techniques yield distances
consistent with the redshift-implied values, i. e., $\sim$ 10 Mpc for the low- 
and $\sim$ 65 Mpc for the high-redshift components.

An independent argument for two distances in SQ involves consistency of
SQ redshifts with those measured in the surrounding field. Galaxies
projected within about one degree of SQ show two preferred redshifts
near cz=  800 and 6500 \kms (Lynds 1972; Materne and Tammann 1974;
Allen and Sullivan 1980 and Shostak et al.  1984). The low redshift
galaxies are shaded in Figure 1a to distinguish them from galaxies with
redshifts near 6500 \kms. Two additional galaxies with redshifts
similar to NGC~7320 were  identified in Shostak et al. (1984) and lie
outside the field shown in Figure 1a.  The redshift data suggest that
the two nearest supercluster structures in the direction of SQ lie at
distances of about 10 and 65 Mpc respectively.  Rejection of the
Doppler interpretation for the high or low redshift parts of SQ would
require nearby galaxies to also show discordant redshifts or a
fortuitous match of discordant redshifts in SQ with cosmological shifts
in the neighboring field.

\section{Past History: Encounters with a Nearest Neighbor NGC~7320C}

We interpret the kernel of three similar redshift members in SQ as a
physical triplet and the core of SQ. These galaxies and NGC~7320C are
hatched in Figure 1b.  The morphological evidence for physical
membership in the triplet is least ambiguous for NGC~7319.  NGC~7317
and 7318A show early-type (E2) morphologies and are therefore less
sensitive to the effects of gravitational encounters. It is somewhat
surprising to find an E-dominated triplet in the field at all, but this
morphology might reflect past secular evolutionary effects within the
group. However, neither of these ellipticals show an unusual color or
signs of geometrical distortion (Schombert et al. 1990; Zepf et al
1991; paper I). 
The strongest evidence for physical membership of NGC~7317
and 18A comes from the diffuse optical light that surrounds them
and that is probably caused by dynamical stripping processes (Arp 1973;
Arp and Lorre 1976; Schombert et al.  1990; paper I). NGC~7318B with
$\Delta$V$\sim -$850 \kms~ relative to the triplet mean velocity is
assumed to be a recent arrival and not relevant for the past dynamical
history of SQ (see next section).

The past dynamical history must account for evidence that points to
past interactions, especially the following: (1) parallel tidal tails (Arp and
Kormendy 1972); (2) stripping of most of the HI from NGC~7319 (Shostak
et al. 1984) and (3) diffuse optical light that surrounds the triplet
kernel. The first two observations are probably related because an
extrapolation backwards of the brighter tail passes  very close to the
projected center of N7319. The parallel extension of the tails and the
fact that both point towards NGC7320C suggest a common origin involving
the galaxy.  Simulations of galaxy interactions do not usually produce
parallel tails (e.g. Howard et al. 1993) leading us to propose that we
may be observing the remnants of two past encounters which suggest that
NGC~7320C may be a loosely bound member of the system. The tails can
reasonably be interpreted as mapping the trajectory of these past
encounters if they were caused by NGC~7320C, an hypothesis that we
favor here.  In the recent past NGC~7320C would have been inside the
group and would have formed an accordant (defined as $\Delta$V$<$10$^3$
km s$^{-1}$from group median) quartet with the triplet kernel. That
accordant quartet would have satisfied the HCG selection criteria only
if NGC~7320 were not superposed. We assume that NGC~7318B had not yet
arrived on the scene.

Naturally the spiral member of the SQ kernel (NGC~7319) shows the most
dramatic evidence for tidal disruption: 1) the spiral arms are
asymmetric, with the eastern arm split into two concentric arcs and the
western one ending in one of the tidal tails; 2) the spiral arms show
no HII regions on H$\alpha$ images that show them in and near NGC~7318A,B
and NGC~7320 (Arp 1973; paper I); 3) NGC~7319 has been stripped of
essentially all HI and 4) it shows a Seyfert nucleus and associated
radio/optical jet (see paper I and Aoki et al. 1996). Figure 2c
shows an H$\alpha$ image (again thanks to W. Keel) centered on NGC7319
that illustrates points 2 and 4. High resolution 21cm observations 
(Shostak et
al. 1984) reveal two extended clouds of HI with redshift similar to the
stable kernel (see Figure 1b).  The total mass of stripped HI with
velocity near 6500 km/s is 1.4$\times$10$^{10}$ h$^{-1}$ M$_{\odot}$ ($ h$ =
H$_o$/100).
This exceed the HI mass of luminous Sb spirals like NGC~7319 by more
than a factor of three and implies that some of the stripped gas may
have come from other galaxies. The proposed intruder, NGC~7320C, shows
a spiral or ringed morphology without evidence for any non-stellar
material. The HI velocity listed for NGC~7320C in RC3 catalog is taken from old
grid mapping with the Arecibo telescope. Higher resolution and
sensitivity observations (Shostak et al.  1984) report no HI  at the
position of NGC~7320C, therefore some of the stripped HI within SQ may
have originated from that galaxy as well.   We conclude that  most of
the damage to NGC~7319 was caused by a direct collision with NGC~7320C
(see also Shostak et al. 1984).

The twin tails and diffuse light in SQ show blue colors consistent with
recent or ongoing star formation. Grid photometry obtained for SQ by
Schombert et al. (1990) find B$-$V $=$ 0.57  for the brighter tail
which is almost as blue as the foreground Sd spiral NGC~7320. The
luminosity of this tail represents about 18\% of the luminosity of
NGC~7319.  The brightest part of the fainter tail was also detected 
and showed colors in the
range 0.1-0.7.  Similarly the diffuse light shows colors similar to the
disks of spiral galaxies B-V= 0.5-0.7. The interstellar matter (ISM) 
 in NGC~7319 and
possibly other components were stripped and heated to $\sim$10$^6$ K (by
analogy to the similar ongoing event involving NGC~7318B, see next
section).  The clouds would then expand, recombine  and cool over a
time-scale $\sim$5$\times$10$^8$ years. Deep images reveal
condensations consistent with the size, color and brightness of HII
regions (or blue clusters) scattered throughout the halo (see e.g.  Arp
and Lorre 1976). The data supports the hypothesis that much of the
stripping occurred in the past few 10$^8$ years and also that some tidally
induced star formation is occurring or has recently occurred, in the
halo and tails.

If the tidal tails map the trajectory of NGC~7320C then it has traveled
$\sim$105 h$^{-1}$ kpc on the plane of the sky since the most recent passage
through SQ. NGC~7320C shows an approximate line of sight velocity 700
\kms~ lower than NGC~7319 (a ratio of apo- to pericenter of 5-6
assuming NGC~7318A is near the group center of mass). Assuming a
transverse velocity equal to the line of sight value suggests that the
collision occurred as recently as 1.5$\times$10$^8$h$^{-1}$ years ago. The
fainter tail is more diffuse (2$\times$ broader than the brighter one)
which is consistent with the idea that it represents an earlier passage
at least t$\geq$ 5$\times$10$^8$ years ago. The roles of NGC~7317 and
7318A in the past dynamical activity or even their past morphologies
are impossible to ascertain. An earlier encounter involving NGC~7320C
and the southwestern-most tail however would provide a mechanism to
account for much of the $\sim$6500 \kms~ HI and blue halo condensations
that are seen west of NGC~7319. The implied trajectory of that tail
passes under NGC~7320 and is consistent with a past encounter involving
NGC~7318A. We cannot rule out the possibility that NGC~7318A is the
remnant bulge of an early spiral whose disk was disrupted in the past
(see paper I). In such a view the high early-type galaxy fraction would
be a product of dynamical evolution in SQ.

\section{Present History: A Penetrating Encounter with NGC~7318B}

The present dynamical history of SQ centers on NGC~7318B which shows an
850 \kms~ radial velocity difference relative to the SQ kernel (these
four galaxies are hatched in Figure 1c). This is so large that one
might argue that this galaxy is unrelated to the group (an accordant
redshift projection) or that it has been dynamically ejected by the
triplet kernel. Two observational clues, however, favor the
interpretation that NGC~7318B is currently colliding with SQ and for
the first time. NGC~7320C is  not actively involved in the
dynamical evolution of the group at this time.

First, we have reanalyzed the lower velocity HI clouds detected 
in SQ. Shostak et al. (1984) found four, spatially-distinct HI 
clouds in SQ with velocities near 5700,
6000 and 6600 (two clouds) \kms.  The latter clouds were discussed in
the previous section. We interpret the two lower velocity clouds as a
single feature. They are adjacent to one another and are centered on
the nucleus of NGC~7318B (indicated schematically in Figure 1c or see
Figure 4 in Shostak et al. 1984 and paper I).
These clouds were previously assumed to be
stripped material like the more massive and extensive $\sim$6500 \kms~
HI clouds. Our reanalysis suggests that they are still associated with
the spiral disk of NGC~7318B.  That galaxy can plausibly be assigned an
Sb or SBb type and therefore would be expected to have a considerable
($\sim$5$\times$10$^9$M$_{\odot}$) HI mass.   They show a velocity gradient
consistent with rotation and a central hole coincident with the nuclear
bulge region of that galaxy (see Figure 6 in the previous reference).
Such holes in the 2D distribution of HI in spiral galaxies with a
prominent population II bulge are often seen. The velocity profile of
these two clouds taken together show a slightly double-horn structure
characteristic of a rotating inclined spiral galaxy. The total HI mass
would be approximately 4.1$\times$10$^9$ h$^{-1}$ M$_{\odot}$ with a FWZI
velocity gradient of 400 km/s consistent with values for spiral
galaxies of similar type and luminosity. The association of this much
neutral gas with NGC~7318B suggests that it still retains a significant
part of its ISM and, therefore has not previously passed through SQ.

The second evidence for a direct collision between NGC~7318B and SQ
involves VLA radio and recent X-ray data of SQ. ROSAT-HRI mapping 
(Pietsch et al. 1997) reveals an elongated structure on the east side
of NGC~7318B partly coincident with one of its spiral arms. This
structure shows a rather close spatial correspondence with a radio
continuum arc (Van der Hulst and Rots 1981). This structure appears to
be sharply bounded and is most easily interpreted as a shock front (it
is indicated in Figure 1c and is also well seen in the H$\alpha$
image shown in Figure 2c).  The origin of the shock would most
plausibly be ascribed to a high velocity
($\Delta$V$\approx$10$^3$ \kms) collision between the ISM in NGC~7318B
and the stripped gas in SQ. The collision must be ongoing and in its
early stages because NGC~7318B retains most of its HI, many HII regions
and a reasonably symmetric spiral pattern. Time limits on
the duration of this event come from two sources:  1) the lifetime of
synchrotron electrons in the radio arc must be a few $\times$ 10$^7$
years without a local source of acceleration (van der Hulst and Rots
1981) and 2) the observed line of sight velocity difference between SQ
and NGC~7318B give a similar timescale for the duration of the intruder
passage through the group.

In our view, NGC~7318B has replaced NGC~7320C as a high velocity
intruder in SQ.  The shock is assumed to be associated with the ongoing
stripping of NGC~7318B.  Optical data also support this view, and the
support becomes even more striking when the image of NGC~7318A is
subtracted (see paper I). Most of the complex structure around
NGC~7318A and B is consistent with spiral structure associated with the
latter high velocity intruder. HII regions can also be seen in the
spiral arms as a further indication that this galaxy has not previously
sustained a major stripping encounter.  The radio/Xray feature
interpreted as a shock front indicates the interface between the
unstripped gas disk of NGC~7318B and the ISM in SQ. The narrowness of
the shock front suggests that the collision is most easily modeled with
NGC~7318B entering the far side of SQ oriented nearly edge-on.

\section{The Implications of SQ as a Typical Compact Group}

The twin problems posed by compact groups center around the large
number of discordant redshift components that they contain, and their
gravitational stability. In SQ, at least, the evidence
favors the chance projection explanation for discordant NGC~7320. The
gravitational stability issue is a problem because attempts at modeling
the groups (e.g. Mamon 1986; Barnes 1989) indicate that they should be
unstable to collapse on short time scales. In this case the number of
compact groups observed today implies that a large merger post-cursor
population should exist. Little or no evidence is found for either the
mergers in progress (Zepf et al 1991; Moles et al 1994) or for the
merger post-cursor population (Sulentic and Raba\c ca 1994). This has
led some to propose that the groups are not real physical systems at
all (e.g. Mamon 1986; Hernquist et al. 1995). The large volume of data
for SQ can shed some light on these problems, particularly, if our
interpretation of its dynamical history is correct {\bf and} if it is
representative of the compact group phenomenon.

Our analysis indicates that SQ is not simply a projection of accordant
redshift galaxies. It is a dynamically active system and evidence for
this activity is virtual proof that it is a physical aggregate.
However the compact group aspect (defined as N$\geq$ 4 accordant
members) is transient in the sense that no fourth tightly bound
component can be identified at this time.  Instead SQ contains a kernel
of three galaxies that must have sufficient mass to attract near
neighbors into high velocity encounters. Our best estimate (paper 1)
for the combined stellar (galaxy+halo), HI and X-ray gas mass is 
M$_{SQ}$$\sim$ 0.75-1.5$\times$10$^{12}$ h$^{-1}$M$_\odot$. Infall
velocities $\sim$800 km s$^{-1}$ were found in n-body simulations of
groups with total mass $\sim$2$\times$10$^{13}$M$_\odot$ (Governato et
al. 1996).   The infalling galaxies result in dynamical activity that is
distinctly episodic. Any catalog of isolated groups will be biased
against many of the groups with a currently infalling intruder because
the intruder will often ``bridge'' the group into its environment and
prevent it from satisfying an isolation criterion. NGC7318B is an
example of a ``safe'' (because it is internal) intruder and NGC~7320C
is one that almost prevented SQ from satisfying the HCG sample selection 
criteria.

Is  SQ a stable virialized system? This question was first directed
towards SQ by Limber and Mathews (1960) before the redshift of NGC~7320
was known. If one repeats this calculation (with M/L$_B$= 13h for
NGC~7317, 18A and 20C as well as M/L$_B$= 8h for NGC~7318B and 19) it
suggests that the triplet kernel is stable with 2T/-$\Omega$$\sim$0.8
but that inclusion of NGC~7318B and 20C yields 2T/-$\Omega$$\sim$ 11.
NGC7320C forms a marginally stable quartet with the triplet kernel
with 2T/-$\Omega$$\sim$ 2. These estimates do not take into account the
large amount of (baryonic) mass present in the halo or the possible
role of non-baryonic matter that might be needed to produce the
observed infall velocities.  The virial calculation and optical images
(the twin tidal tails) interpreted as evidence for recent
passage(s) through the group suggests that NGC~7320C has been recently
captured by the kernel.  Finally, NGC~7318B is almost certainly
entering the group for the first time.  Thus, SQ is probably a bound
triplet that has captured a fourth member (NGC~7320C) within the past
Gyr and is now being visited by another (NGC~7318B).

Governato et al. (1996) have advanced one of the most complete
scenarios for the formation and evolution of compact groups. They
create groups in a critical universe by 1) seeding them with primordial
merger events and 2) growing present day groups with secondary infall
onto these seeds.  Our study of SQ strongly supports the second part of
this model including the conclusion that the infall will be high
velocity and will resist rapid merging. It also supports the
qualitative discussion of Moore et al. (1996) that sees high velocity
intruders as an effective means to dynamically evolve galaxies and
create diffuse halos.  They consider random high velocity encounters in
clusters while compact groups are found in non-cluster environments 
(Sulentic 1987). While high relative velocities in clusters are due to the
high internal velocity dispersion, the lower mass of small groups such as 
SQ implies that only infalling galaxies can achieve velocities V$>$
500 km s$^{-1}$.

In our case the seed is not a single merger remnant as proposed by
Governato et al. (1996) but a triplet. The over-representation of
luminous elliptical galaxies in the triplet suggests that it may also
have experienced strong dynamical evolution as it formed perhaps by a
random initial encounter between three spirals(?). It is not clear if
dynamical evolutionary effects have significantly altered the
properties of NGC~7317 or NGC~7318A because their morphologies, colors
and kinematics appear rather normal. Thus they could be ellipticals or
remnant spiral bulges. Existing images do not allow a distinction
between these possibilities. An extrapolation of the tail under
NGC~7320 is consistent with an encounter trajectory where NGC~7320C
would have passed very near NGC~7318A as recently as $\sim$ 1Gyr ago.
There is sufficient stripped gas to account for three or four spiral
galaxies. The presence of radio continuum
and X-ray emission in the nuclear regions of NGC~7318A may indicate an
active past history. 

A luminous halo surrounds the triplet which indicates that it is a
dynamically evolved physical system. Our conservative estimate of the
halo luminosity (V band: see paper 1) gives M$_{halo}$= -20.9 +
5log(h)$\geq$ M$_*$.  This is almost ten times the luminosity of the
tail created in the most recent passage of NGC7320C.  Neither tail was
included in this estimate but each will increase the halo luminosity by
$\sim$ 10\%.  It is the high velocity intruders that cause SQ to grow
(in galaxy population and halo mass) and that prevents the triplet
kernel from coalescing by injecting kinetic energy into the group.  In
this view SQ must be a relatively young group unless NGC7320C has been
perturbing it for a longer time; otherwise the triplet should have
merged. This would require the triplet to form and dynamically evolve
in  the past 1-2 Gyr.  Assuming that the halo was created by similar
processes  (a tidal tail at a time) and at about the same rate (one
M$_V$= -18.4 + 5log(h) tail per 5$\times$10$^8$ years), would imply an
age of several Gyr for its formation. SQ may not be primordial but the 
stable-kernel part of it is at least 2 Gyr old.

Neither of the elliptical components in the kernel show luminosities or
other properties consistent with having been recent mergers. In this
respect SQ is typical of other compact groups where no evidence for
ongoing merging is seen (Zepf and Whitmore 1991; Moles et al. 1994). 
If our view is correct, then rather than mergers, these ellipticals may be  
the dynamically evolved remnants of spiral galaxies.  SQ is also 
similar to other compact groups in the sense that its optical and FIR 
emission properties indicate a lower level of current star formation 
activity than is observed in pairs (Sulentic and de Mello Raba\c ca 1994; 
Moles 
et al. 1994). SQ suggests that the lack of starburst activity in 
compact groups is due to the
lack of bound gaseous disks in many component galaxies. In SQ a major
stripping event happened in the recent past and another is in
progress.  The gas is either stripped and neutral or shocked and hot. A
large intergroup star forming region is observed in H$\alpha$ emission 
within the group but outside of the galaxies (see Figure 2c). However 
so much of the gas in SQ is stripped or shocked that it is too 
diffuse(cool or hot) to form
large numbers of stars. The FIR emission from SQ is not strong
especially when allowances are made for possible contributions from
NGC7320 and the Seyfert nucleus in NGC7319. The excess FIR emission 
expected from these sources apparently cannot compensate for the 
deficit emission from SQ component galaxies. The transition from a 
normal star forming disk to the shocked
state should be quite sudden given the high velocity of this intruder.
Given the efficiency and quasi-periodic nature of the tidal
perturbation in compact group, it is tempting to ascribe the Seyfert
activity in NGC7319 to the past intruders as well. Whatever gas was not 
stripped may have been rapidly channeled into the nucleus to fuel active 
galactic nucleus (AGN) activity.

SQ suggests that compact groups consist of a tightly bound subsystem
(kernel or seed) plus a loosely bound, or even unbound, population of
infalling neighbors. The most common situation for HCG groups would 
involve a triplet or quartet acting as a seed plus 1 or two intruders. 
Indeed, it is easier to form
a bound pair than a triplet, but given the n$\geq$4 number criterion
used in HCG, triplets would be more often selected because it is easier
to have a 3+1 rather than  2+2 configuration.  The kind of dynamical
encounters do not lead to rapid merging suggested by many models. The
intruders are stripped which reduces their mass and cross section to
frictional effects. At the same time they inject energy into the kernel
which sustains it against collapse.  Some stars form but not as many as
commonly observed in more dynamically  stable pairs. The more frequent
and high impulse events in compact groups may also foster AGN activity
at a higher rate than pairs do. It is perhaps appropriate that the
first compact group discovered more than a century ago provides the
clearest clues to their origin and evolution.


%
%

\clearpage

\figcaption{(a) Schematic of the Stephan's Quintet area. Low redshift
galaxies (V$_0$$\sim$800\kms) are shaded. All other indicated galaxies
have redshifts near $\sim$6000 \kms. Galaxies discussed in the text are
labeled. The dotted region indicates the area shown in Figure 1b.
(b) Map of the immediate environment of SQ. Galaxies involved in the
past dynamical history are shaded. The two optical tails are indicated
along
with the rough extent of HI clouds near 6500 \kms. (c)
Similar map to Figure 1b but shifted to the west by $\sim$30 arcsec.
Galaxies involved in the present dynamical history are shaded. The
region of the radio synchrotron and X-ray shock feature is indicated
along with the approximate extent of HI clouds near
5700 and 6000 \kms.}

\figcaption{(a) R band image of SQ obtained with the  1.5m telescope at 
Calar Alto (Spain).
The large and small boxes
indicate the field  of
view for the low and high redshift H$\alpha$ images shown in 2b and 2c
respectively.
(b) Continuum subtracted low-redshift H$\alpha$ image for NGC~7320 and
projected
tidal
tail. Images were obtained with a 1m telescope at Lowell Observatory
using a
398$\times$398 pixel CCD (0.706 arcsec/pixel). H$\alpha$ image was the
average of four 20 min.  exposures through a filter with effective
width of 70$\rm{\AA}$ centered at 6600$\rm{\AA}$. Continuum frame was a
single 10 min. R exposure. (c) Continuum subtracted high-redshift
H$\alpha$ image for NGC~7318b and 19 and the region of the shock
front.
Images were obtained with the 2.1m telescope at KPNO using a
796$\times$796 pixel CCD (0.194 arcsec/pixel). The H$\alpha$ filter had
a peak wavelength at 6693$\rm{\AA}$ with FWHM of about 65$\rm{\AA}$. It
was a single 20 minute exposure. There was a matching 10 minute R band
continuum exposure. We thank Bill Keel for kindly providing access to
these unpublished images.}


\small
\begin{thebibliography}{}

\bibitem[Allen and Sullivan 1980]{as80} Allen, R.J., and Sullivan III,
W.T.
1980,
\aap, 184, 181

\bibitem[Aoki et al. 1996]{ao96} Aoki, K., Ohtani, H., Yoshida, M., and
Kosugi,
G.
1996, \aj, 111, 140

\bibitem[Arp 1973]{a73} Arp, H. 1973, \apj, 183, 411

\bibitem[Arp and Kormendy 1972]{ako72} Arp, H., and Kormendy, J. 1972,
\apjl,
178, L111

\bibitem[Arp and Lorre 1976]{al76} Arp, H., and Lorre, J. 1976, \apj,
210, 58

\bibitem[Barnes 1989]{ba89} Barnes, J.E. 1989, Nature, 338, 123

\bibitem[Burbidge and Burbidge 1961]{bb61} Burbidge, E.M., and
Burbidge, G.R.
1961,
\apj, 134, 244

\bibitem[Bushouse 1987]{b87} Bushouse, H. 1987, \apj, 320, 49

\bibitem[Governato et al 1996]{g96} Governato, F., Tozzi, P. and
Cavaliere, A. 1996,  \apj, 458, 18

\bibitem[Hernquist et al 1995]{h95} Hernquist, L., Katz, N. and
Weinberg, D. 1995, \apj, 442, 57

\bibitem[Hickson 1982]{h82} Hickson, P. 1982 \apj, 255, 382

\bibitem[Howard et al 1993]{hkb93} Howard, S., Keel, W., Byrd, G. and
Burkey, J., 1993, \apj, 417, 502


\bibitem[Kent 1981]{k81} Kent, S. 1981 \pasp, 93, 554

\bibitem[Limber and Mathews 1960]{lm60} Limber, D. and Mathews, W.
1960, \apj, 132, 286

\bibitem[Lynds 1972]{ly72} Lynds, C.R. 1972, {\it In External Galaxies
and
Quasi-Stellar Objects}, ed. D.S. Evans, (Reidel:Dordrecht), p. 376

\bibitem[Mamon 1986]{ma86} Mamon, G.A. 1986, \apj, 307,426

\bibitem[Materne and Tammann 1974]{mt74} Materne, J. and Tammann, G.
1974, \aap, 35, 441

\bibitem[Moles et al. 1994]{mopmmc94} Moles, M., del Olmo, A., Perea,
J., Masegosa, J., M\'arquez, I., and Costa, V. 1994, \aap, 285, 404

\bibitem[Moles et al. 1997]{mms97}Moles, M., M\'arquez, I., and
Sulentic, J.W., 1997, A \& A submitted, paper I.

\bibitem[Moore et al. 1996]{moo96} Moore, B., Katz, N., Lake, G.,
Dressler, A., and Oemler Jr., A.  1996, Nature, 379, 613


\bibitem[Pietsch et al 1997]{p97} Pietsch, W., Trinchieri, G., Arp, H.
and Sulentic, J.  1997, \aap,
in press

\bibitem[Schombert et al. 1990]{sch90} Schombert, J.M., Wallin, J.F. and
Struck-Marcell, C. 1990, \aap, 99, 497

\bibitem[Shostak et al. 1984]{ssa84} Shostak, G.S., Sullivan III, W.T.,
and
Allen, R.J.
1984, \aap, 139, 15

\bibitem[Sulentic 1987]{sul87} Sulentic, J.W. 1987, \apj, 322, 605

\bibitem[Sulentic 1994]{sul94} Sulentic, J.W. 1994, In {\it Progress in
New Cosmologies}, eds. H. C. Arp et al., (Plenum) p. 49

\bibitem[Sulentic and de Mello Raba\c ca 1994]{smr94} Sulentic, J.W.,
and de
Mello
Raba\c ca, D.F. 1994, \apj, 410, 520

\bibitem[Sulentic and Raba\c ca 1994]{sr94} Sulentic, J.W., 
and Raba\c ca, C.
1994,
\apj, 429, 531

\bibitem[van der Hulst and Rots 1981]{hr81} van der Hulst, J.M., and
Rots, A.H.
1981,
\aj, 86, 1775

\bibitem[Zepf et al. 1991]{zwl91} Zepf, S.E., Whitmore, B.C., and
Levison, H.F.
1991, \apj, 383, 524

\end{thebibliography}
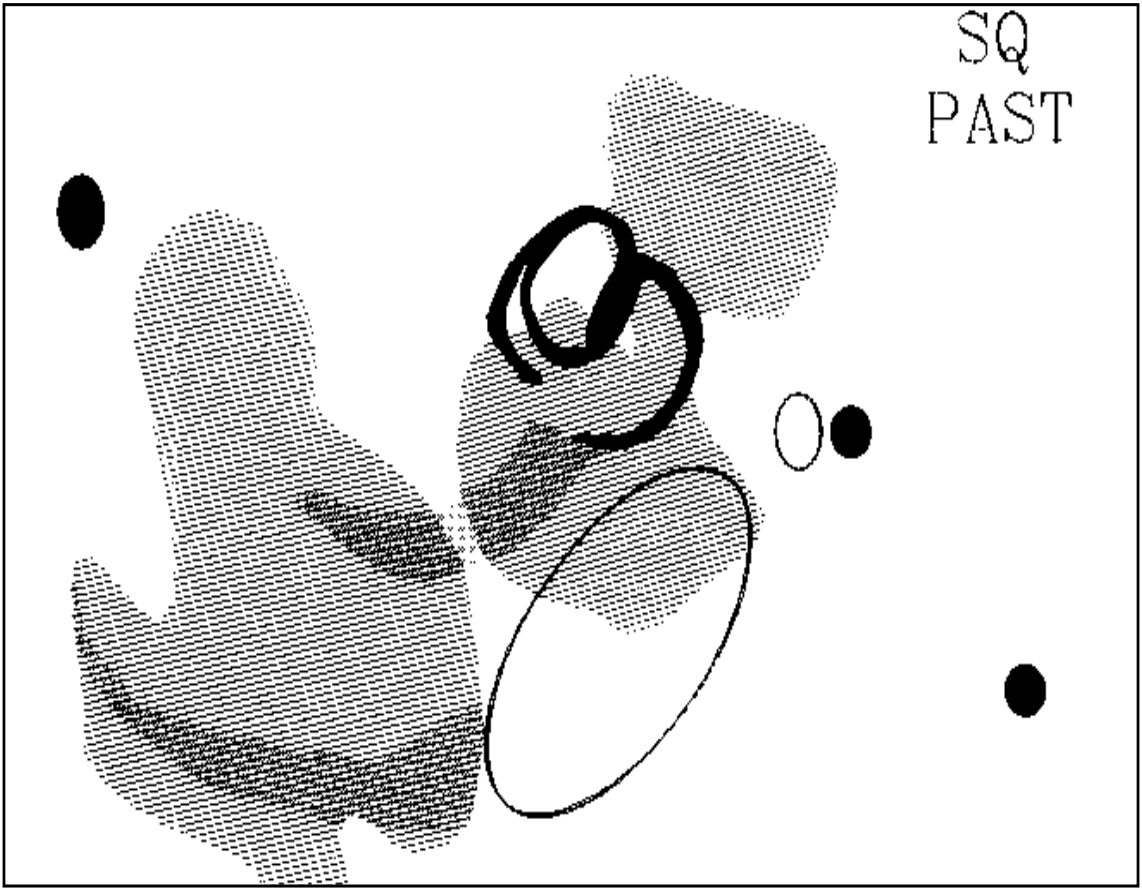
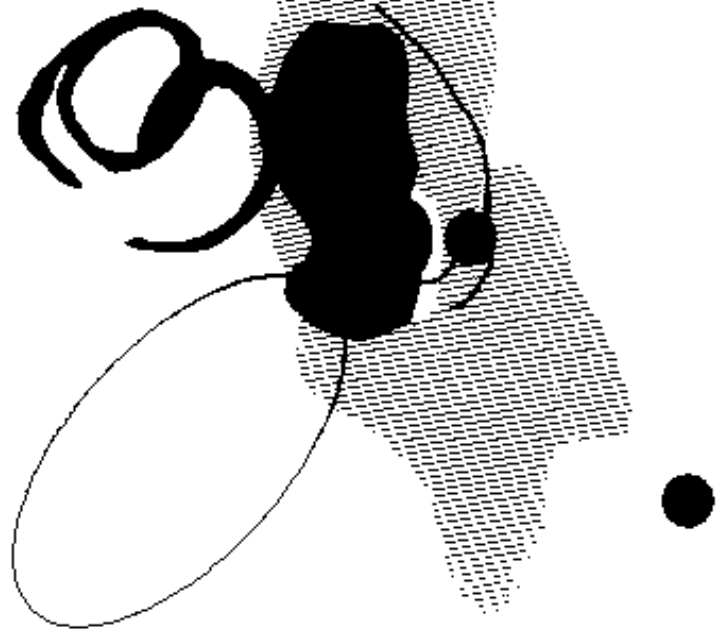
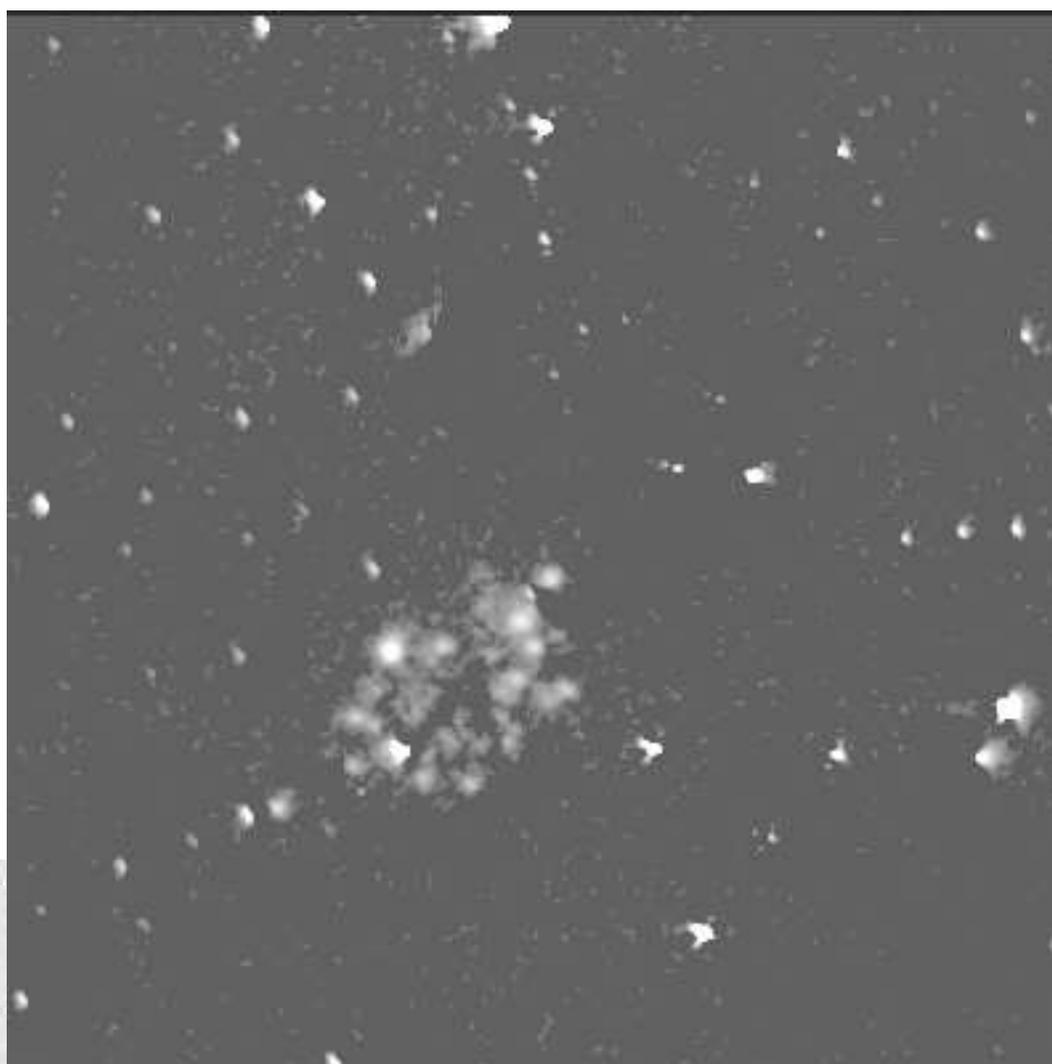
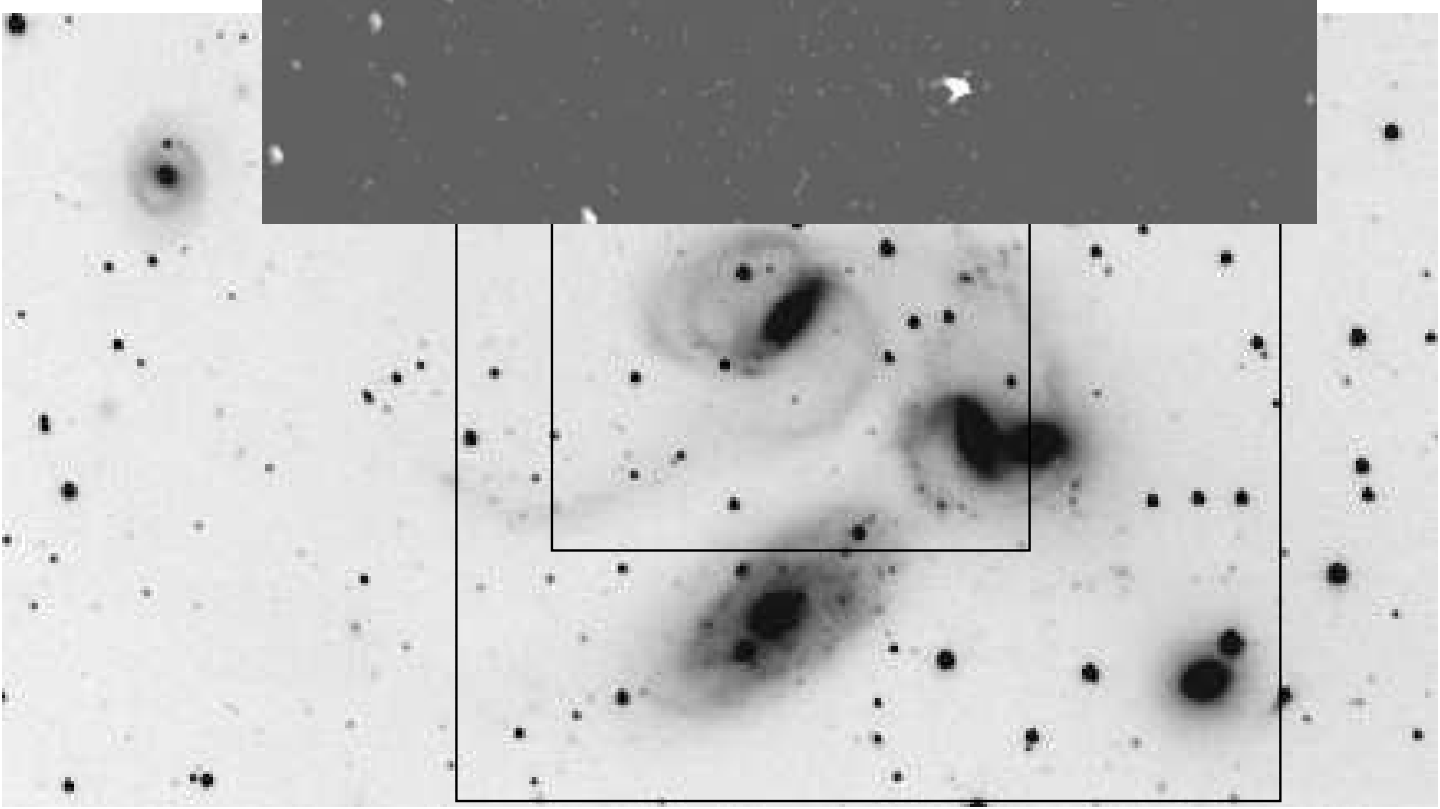
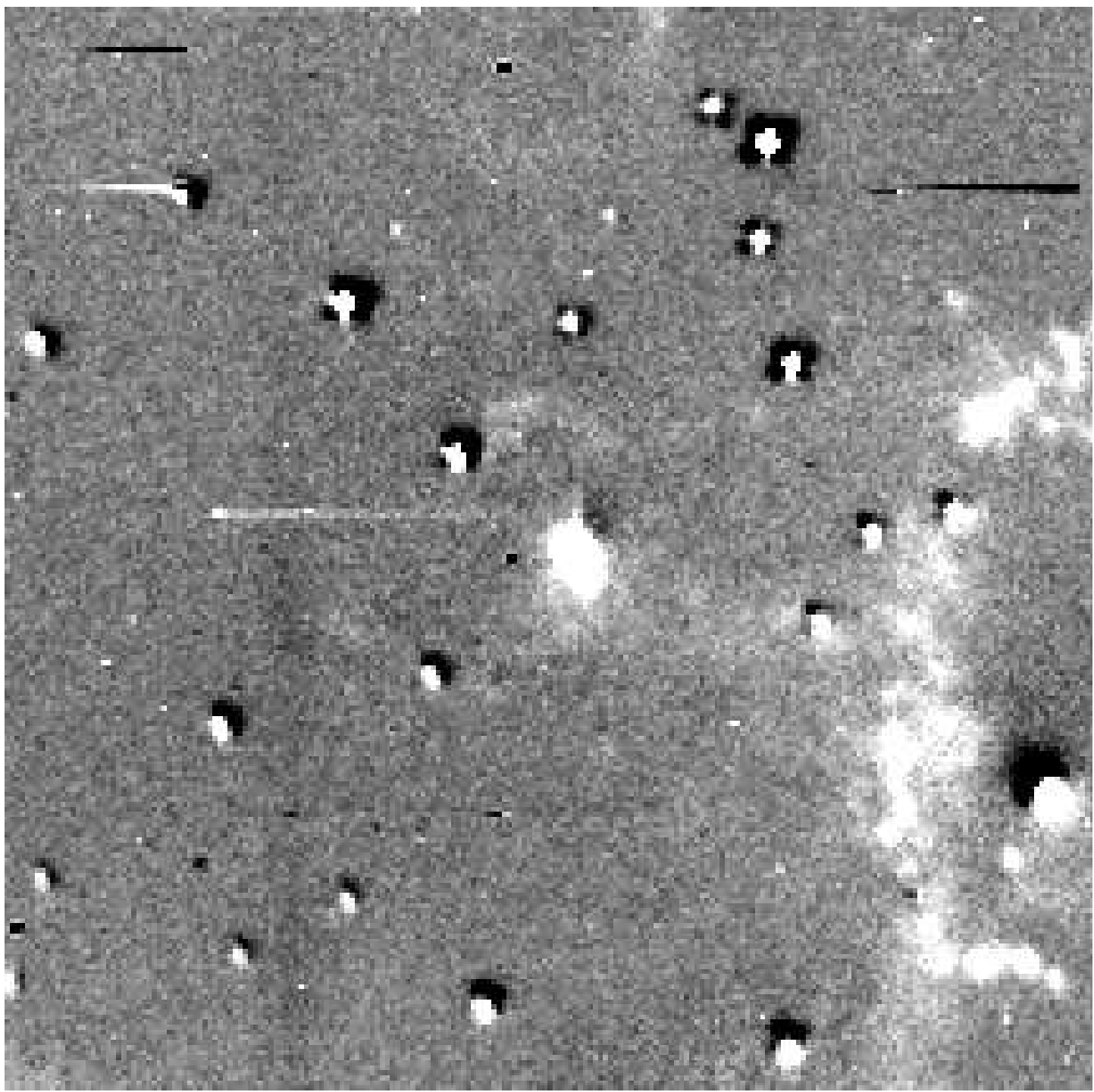
\end{document}